\documentclass[a4paper,11pt]{article}
\usepackage{pos}
\usepackage{wrapfig}
\newcommand{\virgolette}[1]{``#1''}

\title{The GRINTA hard X-ray mission: an Explorer of the Transient Sky}

\manuallySeparateAuthors

\author*[a]{James Rodi}
\author[a]{and Lorenzo Natalucci}
\author[]{on behalf of the GRINTA collaboration}

\affiliation[a]{Istituto di Astrofisica e Planetologia Spaziali,\\
  Via Fosso del Cavaliere, 100, Rome, Italy}

\emailAdd{james.rodi@inaf.it}

\abstract{The era of time domain multi-messenger (MM) astrophysics requires sensitive, large field-of-view (FoV) observatories that are able to quickly react in order to respond to alerts from gravitational wave (GW) triggers, neutrino detections, and transient sources from all parts of the electromagnetic (EM) spectrum.  This is particularly true at hard X-rays and soft gamma-rays where the EM counterparts to GW triggers, gamma-ray bursts (GRBs), emit most of their flux.  While the present decade has a number of instruments capable of accomplishing this task, there are no missions planned for the 2030's when improved MM facilities will detect many more events.  It is in this context that we present the GRINTA mission concept.  GRINTA has a large area, large FoV detector to search for short, impulsive events in the 20 keV - 10 MeV energy range and a coded mask telescope for localizing and performing follow-up observations of sources from 5-200 keV.  While GRINTA's main scientific goal is studying MM events, the instruments will observe numerous other sources to explore the sky at hard X-rays/soft gamma-rays.}

\FullConference{Frontier Research in Astrophysics – IV (FRAPWS2024)\\
 9-14 September 2024\\
Mondello, Palermo, Italy\\}

\renewcommand{\logo}{\relax}

\begin{document}
\maketitle

\section{Introduction}

The year 2017 started the multi-messenger (MM) astrophysics era with the detection of gravitational waves (GWs) coincident with a short gamma-ray burst (SGRB) \citep{2017ApJ...848L..12A} and the association of neutrinos with the flaring blazar TXS 0506+056 \citep{2018Sci...361.1378I}.  These events produced a wealth of results. However, subsequent years have yielded few new events to better understand the processes at work in these sources, particularly with no GW/electromagnetic (EM) counterparts since GW170817.  With the planned upgrades and new GW detectors (LIGO/Virgo/KAGRA and 3rd generation detectors \citep{2020CQGra..37p5003A,2024icrc.confE1591C,2019BAAS...51g..35R,2023JCAP...07..068B}) and neutrino facilities (IceCube-Gen2 and KM3NET 2.0 \citep{2024icrc.confE.994I,2016JPhG...43h4001A}), the decade of the 2030's is primed for breakthrough discoveries in MM astrophysics.  Nonetheless, a key energy range is missing to maximize the results from these observatories.  To date there are no accepted hard X-ray/soft gamma-ray missions scheduled to fly in the 2030's.  This energy band is critical for detecting, localizing, and studying the physical processes in MM events, especially GW events accompanied by GRBs.  It is in this context that we present the mission concept GRINTA (Gamma-Ray INternational Transient Array).  The payload is composed of a hard X-ray imager (HXI, 5-200 keV), and a large field-of-view (FoV) transient event detector (TED, 20 keV - 10 MeV) that together with a rapid, autonomous re-pointing spacecraft are capable of detecting, localizing, and characterizing both the prompt and afterglow emission from GRBs and other transient phenomena.  When operational, GRINTA will work in synergy and coordination  with other facilities planned for the 2030's across the EM spectrum (e.g. SKA, LSST, CTA).

\section{Science Case}

\subsection{Gravitational Wave and Their EM Counterparts}

To date no firm EM counterparts have been detected for a binary black hole (BBH) merger, despite numerous GW detections, which suggests that typically they are not EM sources \citep{2016AAS...22820803B,2016ApJ...821L..18P,2016ApJ...819L..21L}.  However, the binary neutron star (BNS) merger GW 170817 and its EM counterpart (GRB 170817A) demonstrated that such events are a SGRB progenitor system, in addition to the host of discoveries garnered across the EM spectrum from follow-up observations \citep{2017ApJ...848L..12A}.  While results from the follow-up observations are important, the localization and characterization of the prompt emission at hard X-ray/soft gamma-ray energies is crucial to the subsequent analyses, as well as providing additional information not available at different wavelengths.  

GW 170817/GRB 170817A provides an example of what can be learned from future events and what questions remain.  The arrival time of the GW and EM signals had a difference of \(\sim 1.7\) s \citep{2017ApJ...848L..13A}.  This difference combined with the distance provided by the GW signal enabled investigations of fundamental physics, like the speed of gravity, Lorentz invariance limits and the equivalence principle \citep{2017ApJ...848L..13A}.  But the reason for the time difference and how consistent the time difference is between GRBs requires observations of more events to further understand.

GBM found the observational characteristics of the prompt gamma-ray emission of GRB 170817A were similar to other SGRBs \citep{2017ApJ...848L..14G}, but its luminosity was extremely low \citep{2017ApJ...848L..13A}.  The reason for the low luminosity is still unclear.  The two most likely explanations are a relativistic jet viewed off axis or a mildly relativistic shock breakout \citep{Gottlieb2018}.  Thus it is unclear how different GRB 170817A is from other SGRBs without additional joint GW/EM detections.  Additionally, NS-BH mergers are potentially a SGRB progenitor system.  Therefore, the relative number of BNS to NS-BH mergers requires more events to better understand the progenitor population.  

The remnant of the BNS merger is another unresolved question.  Three scenarios have been proposed: the direct formation of a BH, the formation of a short lived NS (lasted \(\sim 100\) ms to hours) before forming a BH, or the formation of a stable NS \citep{2019MNRAS.486.4479L}.  No extended emission was found after GRB 170817A \citep{2017ApJ...848L..14G,2017ApJ...848L..15S}, which suggests that a stable NS was not formed.  However, the first pointed X-ray observations began \(\sim 20\) hr post merger \citep{2017ApJ...848L..15S} so it was not possible to search for a short-lived NS.  Thus GW~170817/GRB 170817A was a groundbreaking event, which provided a wealth of information about fundamental physics and and astrophysics, but there are still numerous questions unanswered by this one event and also new questions about how common such an event is.  The answers to these questions (and others) require more joint GW/EM detections. GRINTA observations in the 2030's in conjunction with improved GW detectors will be important to uncovering the answers to these unresolved issues.

\subsection{Neutrino Events}

Another area of MM investigation is sources of high-energy neutrinos.  These weakly interacting particles are able to travel long distances, and thus can provide complementary information to photons.  Currently, only a few sources have been associated with neutrinos (see \citep{2024arXiv240406867K} and references within), though the associations are only marginally significant.  The best known is the blazar TXS 0506+056, from which neutrinos were detected during a weeks-long gamma-ray flare \citep{2018Sci...361.1378I}, implying that the neutrinos were produced in the blazar jet.  Recent results report a correlation between the hard X-ray flux and neutrino fluxes for six nearby active galactic nuclei (AGNs) \citep{2024arXiv240406867K}.  Because not all of these AGN have jets, the authors interpret the correlation as evidence that the neutrinos are produced in the AGN cores.  

Other sources of high-energy neutrinos are potentially tidal disruption events \citep{2021arXiv211109390R,2021NatAs...5..510S} and GRBs.  Neutrinos were detected from the core collapse supernova SN1987A \citep{1987ApJ...318L..63B}, which are a progenitor system for long GRBS (LGRBs) \citep{2006Natur.441..463F}.  Also extended emission in SGRBs has been proposed as a source of neutrinos \citep{2022PhRvL.128v1101R}.  Thus hard X-rays/soft gamma-rays combined with next generation neutrino facilities can reveal where and how these particles are produced in numerous classes of objects.  

\subsection{Non-MM Transients}

While searching for EM counterparts to MM events, GRINTA will monitor the sky for other types of transient events.  Recent results found that a subpopulation of SGRBs are actually giant flares from magnetars that are at extra-galactic distances \citep{2021ApJ...907L..28B}.  These events are typically shorter in duration and less luminous than the rest of the SGRB population.  Spectra from these events vary significantly on ms timescales with photons up to at least a few MeV \citep{2021Natur.589..211S,2021Natur.589..207R}.  GRINTA's large FoV, sensitivity, and high time resolution will enable studies of temporal and spectral evolution of magnetar giant flares as well as other types of GRBs.   

Galactic transients are another set of sources that GRINTA will study.  With its broad energy range and large FoV, it can observe sources from the beginning of their outbursts to understand how they trigger, as well as spectral state changes throughout the bursts as the source returns to quiescence.  Additionally, HXI's high point source location accuracy (20 arcsec) allows for follow-up observations across the EM spectrum and for multi-wavelength studies of newly discovered sources.  

Finally, GRINTA will monitor a plethora of AGN from soft X-rays to soft gamma-rays, yielding information about the physical conditions close to the supermassive BH.  

\subsection{X-ray and Gamma-ray Surveys}

\begin{wrapfigure}{r}{0.5\textwidth}
    \centering
    \includegraphics[scale=0.45, angle=0,trim = 10mm 155mm 0.5mm 0mm, clip]{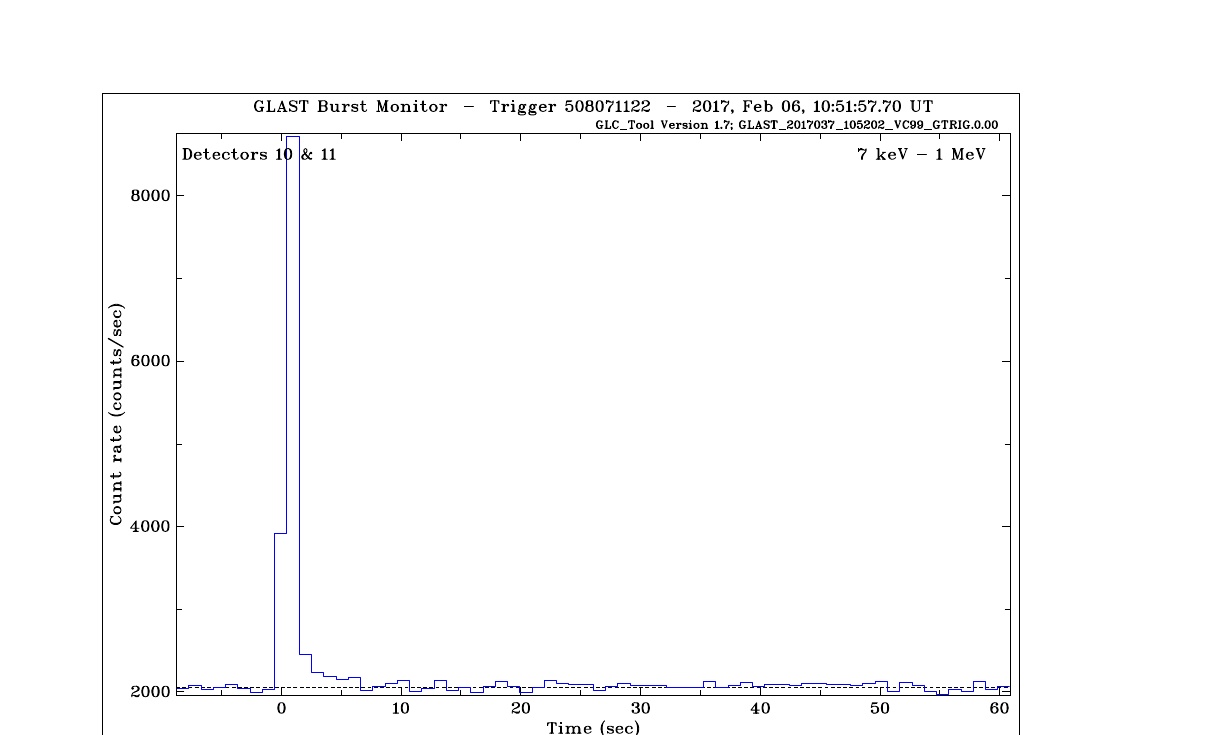}
  \caption{Fermi/GBM light curve of GRB 170206A.}
    \label{fig:grb_example}
\end{wrapfigure}

The INTEGRAL and Swift satellites have been observing the hard X-ray sky for roughly two decades each.  INTEGRAL has focused on the galactic plane, while Swift predominately monitors extra-galactic objects.  Approximately 40\% of IBIS sources and 20\% of Swift/BAT sources do not have an optical or infrared counterpart \citep{2016ApJS..223...15B,2016MNRAS.460...19M,2010A&A...524A..64C,2013ApJS..207...19B,2018ApJS..235....4O}.  Most are likely highly absorbed AGN or galactic X-ray binaries with a NS or a BH.  Also, the eRosita survey has \(\sim 28,000\) soft X-ray sources \citep{2021arXiv210614517B}.

HXI's significant improvement in point source location accuracy and lower energy threshold, relative to INTEGRAL and Swift/BAT, will be able to better localize sources for optical and infrared follow-up observations, as well as search for high-energy emission from eRosita sources.

\section{Instruments}

\subsection{The Transient Event Detector}

\begin{wrapfigure}{r}{0.45\textwidth}
 \centering
  \vspace{-1mm}
    \includegraphics[scale=0.3, angle=0,trim = 0mm 5mm 0.5mm 0mm, clip]{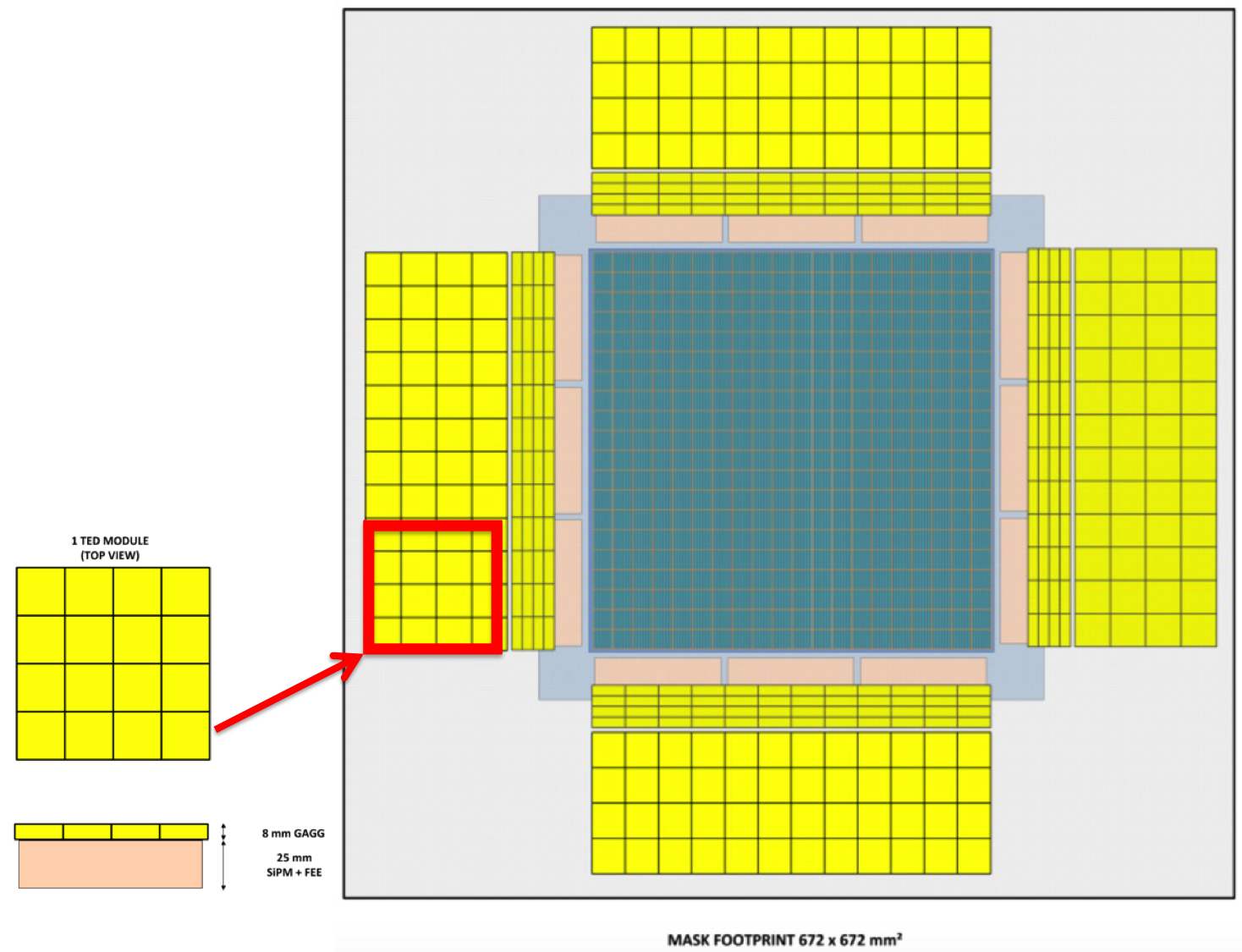}
  \caption{Top down view of GRINTA with TED shown in yellow and HXI in the center.  An example of a TED module is highlighted in red.}
    \vspace{-3mm}
    \label{fig:grinta_top}
\end{wrapfigure}

TED consists of 24 GaGG (Gadolinium Aluminium Gallium Garnet) scintillator modules arranged to view different locations on the sky in the 20 keV - 10 MeV energy range.  Each module has an area of 100 cm$^2$ (a total of 2400 cm\(^2\)) and is attached to a Silicon photomultiplier (SiPM) readout.  This configuration enables observations of \(\sim 8\) sr of the sky in search of short, impulsive events by monitoring the modules' ratemeters for significant excesses above the background count rate.  An example of this can be seen in the light curve of a GRB detected by Fermi/GBM in Figure~\ref{fig:grb_example}.  Events can be localized onboard by comparing the relative count rates between all 24 modules.  We anticipate localizing roughly 80\% of events to \( < 10^{\circ}\).

The 24 GaGG modules are evenly divided over four~\virgolette{walls} around HXI. 
Each wall has two panels of three modules.  One panel is oriented at \(45^{\circ}\) with respect to the HXI pointing direction while the other panel is oriented at \(90^{\circ}\).  See Figure~\ref{fig:grinta_top}, which shows a top down view of GRINTA with HXI in the center surrounded by TED on the sides.  These orientations maximize the exposed detector area, enabling high sensitivity to events far off axis relative to the HXI pointing direction.   Based on Fermi/GBM's onboard sensitivity \citep{2009ApJ...702..791M} and TED's large FoV, we anticipate roughly 90 SGRBs and 480 LGRBs/year.  The joint GW+EM detection rate relies heavily on which GW detectors are operational during the GRINTA mission.  Starting in the mid-2030's the 3rd generation GW detectors Einstein Telescope (ET) and Cosmic Explorer (CE) are expected to be operating.  In the case of ET or ET+CE, the predicted numbers of joint detections with GRINTA are roughly 60 and 85/yr, respectively, following the method outlined in \cite{2023A&A...675A.117R}.

\subsection{The Hard X-ray Imager}

HXI is a coded mask telescope to operate in the 5-200 keV energy range.  The detector plane is an array of \(16 \times 16\) CdTe crystals arranged in a \(16 \times 16\) array (65536 pixels). It is based on a customized version of the Caliste module design \citep{2014JInst...9C5019M}.  Each pixel is 1 mm \(\times\) 1 mm \(\times\) 1.5 mm for a total detection area of 655 cm\(^2\).  The coded mask is made of tungsten with an open fraction of \(1/2\) and 1.8 m above the detector plane.  Thus the fully-coded FoV is \(10^{\circ} \times 10^{\circ}\) (\(29^{\circ} \times 29^{\circ}\) zero response) with an angular resolution of 3.8 arcmin and a source location accuracy of 20 arcsec (10 \( \sigma\)).  The expected 3-\(\sigma\) sensitivity is \(< 6 \times 10^{-3}\) ph/cm\(^2\)/s (5 mCrab) in 10\(^4\) s in the 5-200 keV energy range.  

HXI's fully-coded FoV is selected to match the location accuracy of TED.  Thus when combined with rapid re-pointing capabilities, the spacecraft can autonomously slew to the location of an event detected by TED to better localize it with HXI and perform follow-up observations.  The expected spacecraft slew rate would be approximately 50\(^{\circ}\)/min, which allows for rapid follow-up observations for nearly all TED triggers.  These capabilities are critical in detecting and studying GRBs afterglows.  

\begin{wrapfigure}{r}{0.65\textwidth}
 \centering
    \includegraphics[scale=0.25, angle=0,trim = 0mm 5mm 0.5mm 0mm, clip]{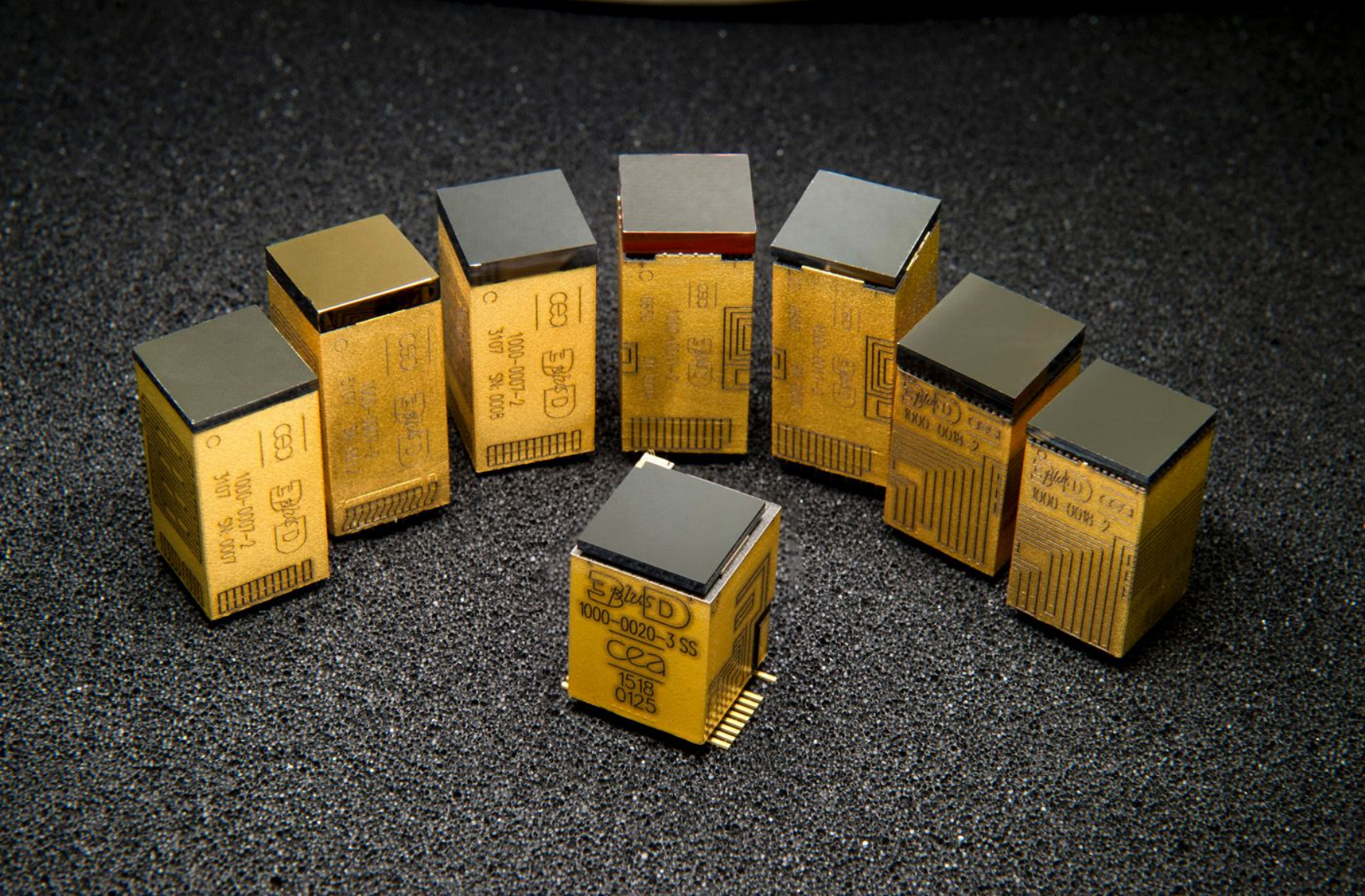}
  \caption{HXI detection unit.}
    \label{fig:grinta_top}
\end{wrapfigure}

For joint GW/TED detections, roughly 65\% are predicted to have detectable afterglows if follow-up observations begin within 60s and approximately 42\% if observations begin within 600s using the same methodology as \citep{2023A&A...675A.117R}.  Additionally, LIGO/Virgo/KAGRA's Post O5 and ET+CE localizations are expected to provide significantly better localizations with a majority of events localized to \(<\) 100 deg\(^2\) \citep{2023JCAP...07..068B}, which matches the HXI FoV.  Thus HXI will be able to perform follow-up observations of the entire uncertainty region to search for EM counterparts.  Assuming a response time of 600s between the GW trigger and the beginning of the HXI observations, the predicted number of joint detections is \(\sim 15\)/yr in the pessimistic case that GRINTA flies before ET and CE are operational, based on \cite{2023A&A...675A.117R}.  Thus in the most optimistic case, the number of joint detections with ET+CE alerts would be considerably higher.


\section{Conclusions}

The next decade is set to make breakthroughs with significant improvements in sensitivities for MM facilities (LIGO/Virgo upgrades, ET, CE, IceCube-Gen2, KM3NET) and EM telescopes (SKA, LSST, CTA, and others).  However, there are no hard X-ray/soft gamma-ray missions planned to fly during that period.  Since many of the sources studied with these observatories emit at all wavelengths, high sensitivity hard X-ray/soft gamma-ray capabilities are needed to maximize the scientific returns.  GRINTA's TED and HXI large FoV, sensitivity, point source location accuracy, and fast re-pointing are well suited fill this gap to find and follow-up EM counterparts of MM events autonomously and alert the community for multi-wavelength observations.  As discussed above, these same attributes are important for studying non-MM events and working synergistically with other EM instruments.


\begin{thebibliography}{99}



\bibitem[Abbott et al.(2017a)]{2017ApJ...848L..12A} Abbott, B.~P., Abbott, R., Abbott, T.~D., et al.\ 2017, ApJL, 848, L12. doi:10.3847/2041-8213/aa91c9

\bibitem[IceCube Collaboration et al.(2018)]{2018Sci...361.1378I} IceCube Collaboration, Aartsen, M.~G., Ackermann, M., et al.\ 2018, Science, 361, eaat1378. doi:10.1126/science.aat1378

\bibitem[Coccia \& Einstein Telescope Collaboration(2024)]{2024icrc.confE1591C} Coccia, E. \& Einstein Telescope Collaboration\ 2024, 38th International Cosmic Ray Conference, 1591

\bibitem[Reitze et al.(2019)]{2019BAAS...51g..35R} Reitze, D., Adhikari, R.~X., Ballmer, S., et al.\ 2019, BAAS, 51, 35. doi:10.48550/arXiv.1907.04833


\bibitem[Adhikari et al.(2020)]{2020CQGra..37p5003A} Adhikari, R.~X., Arai, K., Brooks, A.~F., et al.\ 2020, Classical and Quantum Gravity, 37, 165003. doi:10.1088/1361-6382/ab9143

\bibitem[Branchesi et al.(2023)]{2023JCAP...07..068B} Branchesi, M., Maggiore, M., Alonso, D., et al.\ 2023, JCAP, 2023, 068. doi:10.1088/1475-7516/2023/07/068

\bibitem[IceCube-Gen2 et al.(2024)]{2024icrc.confE.994I} IceCube-Gen2, Abbasi, R., Ackermann, M., et al.\ 2024, 38th International Cosmic Ray Conference, 994

\bibitem[Adri{\'a}n-Mart{\'\i}nez et al.(2016)]{2016JPhG...43h4001A} Adri{\'a}n-Mart{\'\i}nez, S., Ageron, M., Aharonian, F., et al.\ 2016, Journal of Physics G Nuclear Physics, 43, 084001. doi:10.1088/0954-3899/43/8/084001


\bibitem[Bartos(2016)]{2016AAS...22820803B} Bartos, I.\ 2016, AAS

\bibitem[Perna et al.(2016)]{2016ApJ...821L..18P} Perna, R., Lazzati, D., \& Giacomazzo, B.\ 2016, ApJL, 821, L18. doi:10.3847/2041-8205/821/1/L18

\bibitem[Loeb(2016)]{2016ApJ...819L..21L} Loeb, A.\ 2016, ApJL, 819, L21. doi:10.3847/2041-8205/819/2/L21


\bibitem[Abbott et al.(2017b)]{2017ApJ...848L..13A} Abbott, B.~P., Abbott, R., Abbott, T.~D., et al.\ 2017, ApJL, 848, L13. doi:10.3847/2041-8213/aa920c

\bibitem[Goldstein et al.(2017)]{2017ApJ...848L..14G} Goldstein, A., Veres, P., Burns, E., et al.\ 2017, ApJL, 848, L14. doi:10.3847/2041-8213/aa8f41

\bibitem[Gottlieb et al.(2018)]{Gottlieb2018} Gottlieb, O., Nakar, E., Piran, T., et al.\ 2018, MNRAS, 479, 588. doi:10.1093/mnras/sty1462

\bibitem[L{\"u} et al.(2019)]{2019MNRAS.486.4479L} L{\"u}, H.-J., Shen, J., Lan, L., et al.\ 2019, MNRAS, 486, 4479. doi:10.1093/mnras/stz1155

\bibitem[Savchenko et al.(2017)]{2017ApJ...848L..15S} Savchenko, V., Ferrigno, C., Kuulkers, E., et al.\ 2017, ApJL, 848, L15. doi:10.3847/2041-8213/aa8f94

\bibitem[Kun et al.(2024)]{2024arXiv240406867K} Kun, E., Bartos, I., Becker Tjus, J., et al.\ 2024, arXiv:2404.06867. doi:10.48550/arXiv.2404.06867

\bibitem[Reusch et al.(2021)]{2021arXiv211109390R} Reusch, S., Stein, R., Kowalski, M., et al.\ 2021, arXiv:2111.09390

\bibitem[Stein et al.(2021)]{2021NatAs...5..510S} Stein, R., Velzen, S. van ., Kowalski, M., et al.\ 2021, Nature Astronomy, 5, 510. doi:10.1038/s41550-020-01295-8

\bibitem[Burrows \& Lattimer(1987)]{1987ApJ...318L..63B} Burrows, A. \& Lattimer, J.~M.\ 1987, ApJL, 318, L63. doi:10.1086/184938

\bibitem[Fruchter et al.(2006)]{2006Natur.441..463F} Fruchter, A.~S., Levan, A.~J., Strolger, L., et al.\ 2006, Nature, 441, 463. doi:10.1038/nature04787

\bibitem[Reusch et al.(2022)]{2022PhRvL.128v1101R} Reusch, S., Stein, R., Kowalski, M., et al.\ 2022, PhRvL, 128, 221101. doi:10.1103/PhysRevLett.128.221101

\bibitem[Burns et al.(2021)]{2021ApJ...907L..28B} Burns, E., Svinkin, D., Hurley, K., et al.\ 2021, ApJL, 907, L28. doi:10.3847/2041-8213/abd8c8

\bibitem[Svinkin et al.(2021)]{2021Natur.589..211S} Svinkin, D., Frederiks, D., Hurley, K., et al.\ 2021, Nature, 589, 211. doi:10.1038/s41586-020-03076-9

\bibitem[Roberts et al.(2021)]{2021Natur.589..207R} Roberts, O.~J., Veres, P., Baring, M.~G., et al.\ 2021, Nature, 589, 207. doi:10.1038/s41586-020-03077-8

\bibitem[Bird et al.(2016)]{2016ApJS..223...15B} Bird, A.~J., Bazzano, A., Malizia, A., et al.\ 2016, ApJS, 223, 15. doi:10.3847/0067-0049/223/1/15

\bibitem[Malizia et al.(2016)]{2016MNRAS.460...19M} Malizia, A., Landi, R., Molina, M., et al.\ 2016, MNRAS, 460, 19. doi:10.1093/mnras/stw972

\bibitem[Cusumano et al.(2010)]{2010A&A...524A..64C} Cusumano, G., La Parola, V., Segreto, A., et al.\ 2010, A\&A, 524, A64. doi:10.1051/0004-6361/201015249

\bibitem[Baumgartner et al.(2013)]{2013ApJS..207...19B} Baumgartner, W.~H., Tueller, J., Markwardt, C.~B., et al.\ 2013, ApJS, 207, 19. doi:10.1088/0067-0049/207/2/19

\bibitem[Oh et al.(2018)]{2018ApJS..235....4O} Oh, K., Koss, M., Markwardt, C.~B., et al.\ 2018, ApJS, 235, 4. doi:10.3847/1538-4365/aaa7fd

\bibitem[Brunner et al.(2021)]{2021arXiv210614517B} Brunner, H., Liu, T., Lamer, G., et al.\ 2021, arXiv:2106.14517

\bibitem[Meegan et al.(2009)]{2009ApJ...702..791M} Meegan, C., Lichti, G., Bhat, P.~N., et al.\ 2009, ApJ, 702, 791. doi:10.1088/0004-637X/702/1/791


\bibitem[Ronchini et al.(2023)]{2023A&A...675A.117R} Ronchini, S., Stratta, G., Rossi, A., et al.\ 2023, A\&A, 675, A117. doi:10.1051/0004-6361/202245348

\bibitem[Meuris(2014)]{2014JInst...9C5019M} Meuris, A.\ 2014, Journal of Instrumentation, 9, C05019. doi:10.1088/1748-0221/9/05/C05019






\end{thebibliography}
\end{document}